\documentclass[useAMS,usenatbib,usegraphicx]{mn2e}
\usepackage{amssymb}

\title[Disc formation in turbulent cloud cores]{Disc formation in turbulent cloud cores:\\ is magnetic flux loss necessary to stop the magnetic braking catastrophe or not?}
\author[R. Santos-Lima, E. M. de Gouveia Dal Pino and A. Lazarian]{R. Santos-Lima$^{1}$\thanks{E-mail: rlima@astro.iag.usp.br}, E. M. de Gouveia Dal Pino$^{1}$\thanks{E-mail:
dalpino@astro.iag.usp.br} and A. Lazarian$^{2}$\thanks{E-mail: alazarian@facstaff.wisc.edu}\\
$^{1}$Instituto de Astronomia, Geof\'isica e Ci\^encias Atmosf\'ericas, Universidade de S\~ao Paulo, S\~ao Paulo, SP 05508-090, Brazil\\
$^{2}$Department of Astronomy, University of Wisconsin, Madison, WI 53706, USA}
\begin{document}



\maketitle

\label{firstpage}

\begin{abstract}
Recent numerical analysis of Keplerian disk formation in  turbulent, magnetized cloud cores by \citet{santos-lima_etal_2012} demonstrated that  reconnection diffusion is an efficient  process  to remove  the magnetic flux excess during the build up of a rotationally supported disk. 
This process  is induced by fast reconnection of the magnetic fields in a turbulent flow.   In a similar  numerical study,  \citet{seifried_etal_2012} concluded that reconnection diffusion or any other non-ideal MHD effects would not be necessary and turbulence shear alone would provide a natural way to build up a rotating disk without requiring magnetic flux loss. Their conclusion was based  on the  fact that   the mean mass-to-flux ratio ($\mu$) evaluated over a spherical region   with a radius much larger than the disk  is nearly constant in their models.  
In this letter we compare the two sets of simulations  and show that  this averaging over large scales can mask significant real increases of $\mu$ in the inner regions where the disk is built up.  
We demonstrate that turbulence-induced reconnection diffusion of the  magnetic field happens in the initial stages of the disk formation  in the turbulent envelope material that is accreting.
Our analysis  is suggestive that reconnection diffusion is present in both sets of simulations and provides a simple
solution for the ``magnetic braking catastrophe''  which is discussed in
the literature in relation to the formation of protostellar accretion disks.
\end{abstract}


\section{Introduction}

Star formation theory for decades developed under the assumption that magnetic flux is frozen in highly conducting interstellar gas, unless a process of ambipolar diffusion carries neutrals across the magnetic field lines \citep{mestel_spitzer_1956, nakano_tademaru_1972, mouschovias_1976, shu_1983, lizano_shu_1989, li_etal_2008, fatuzzo_adams_2002, zweibel_2002}.
The flux freezing is, however, violated in turbulent fluids due to fast reconnection \citet{lazarian_vishniac_1999} (henceforth LV99). This prediction is now not only supported by successful numerical tests \citep{kowal_etal_2009, kowal_etal_2012a} and observations \citep{ciaravella_raymond_2008, sych_etal_2009}, but also by formal mathematical derivations based on modern understanding of the Lagrangian properties of MHD turbulence (\citealt{eyink_2011, eyink_etal_2011}). The diffusion of magnetic fields mediated by reconnection in turbulent fluids was predicted to be important for star formation by Lazarian (2005).  The process should be widely spread as turbulence is ubiquitous in the interstellar media (Armstrong et al. 1994,\citealt{elmegreen_scalo_2004, mckee_ostriker_2007}, Chepurnov \& Lazarian 2010). The corresponding process was termed {\it reconnection diffusion} in analogy with the ambipolar diffusion, the accepted process in the standard picture for magnetic flux removal from molecular cloud cores. According to the latter mechanism, the violation of magnetic flux freezing is possible through the drift of neutrals passing through nearly
perfectly frozen-in ions with the magnetic field.  The reconnection diffusion presents a potent alternative to the ambipolar diffusion process (see \citealt{lazarian_2011, degouveiadalpino_etal_2012} for reviews).

The numerical demonstration of the reconnection diffusion was first performed by \citet{santos-lima_etal_2010} for the case of  diffuse media and collapsing molecular clouds (see also \citealt{leao_etal_2012, lazarian_etal_2010, degouveiadalpino_etal_2011}).

Recently \citet{santos-lima_etal_2012} (hereafter SGL12)  have performed simulations of the formation of a Keplerian disc  in a turbulent, strongly magnetized low-mass cloud core (with a sink particle in the center with M $= 0.5 $ solar-mass and an initial uniform magnetic field in the z direction with intensity 35 $\mu$G) and found that the reconnection diffusion was efficient in this set up, decreasing the magnetic flux during the disk build up. SGL12 proposed that this mechanism can provide the solution of the so-called ``magnetic braking catastrophe'' discussed in the literature \citep{allen_etal_2003, galli_etal_2006, price_bate_2007, hennebelle_fromang_2008, mellon_li_2008, krasnopolsky_etal_2011}.

The latter conclusion was criticized in \citet{seifried_etal_2012} (hereafter S+12) who performed AMR simulations of the collapse of a 100 solar mass turbulent cloud core permeated by a  magnetic field (with 1.3 mG in the center and declining radially
outwards with R$^{-0.75}$). They introduced sink particles in the cloud above a
density threshold and detected the formation of several protostars around
which Keplerian discs with typical sizes of up to 100 AU built up. Then, they
examined a few mechanisms that could be potentially responsible for lowering
the magnetic braking efficiency and thus, allowing for the formation of the
Keplerian discs and concluded that none was necessary in their models, nor
even reconnection diffusion. They argued that the build up of the Keplerian
disk was a consequence of the shear flow generated by the turbulent motions in the surroundings of the disk (which carry large amounts of
angular momentum). The lack of coherent rotation in the turbulent velocity
field would not allow the development of toroidal B components and toroidal
Alfv\'en waves that could remove outward magnetic flux, in spite of the small
values of the mass-to-flux ratio that they considered in their models (around $\mu \simeq 2-3$).

According to S+12, any effects like misaligned magnetic fields and angular
momentum vectors, reconnection diffusion or any other non-ideal MHD effects seem not to be necessary. They conclude that ``turbulence alone provides a natural and at the same time very simple mechanism to solve the magnetic braking catastrophe''.

S+12 conclusion above was based on the calculation of  the mean mass-to-flux ratio within a sphere around the disk with a radius much larger than the disk  (i.e., r = 500 AU).  This ratio $\mu$ was computed taking the volume-weighted, mean magnetic field  evaluated  in this  sphere, in combination with the sphere mass M, normalized by the critical value. They found  that  at these  scales $\mu$ varies around  a  mean  of  2  -  3 and is comparable with the initial value in the core (which is also  the overall  initial value  in  the massive cloud) and also to the MHD simulations without turbulence.
However, they have also found that in some cases $\mu$ increases at smaller
radius and eventually reaches values above 10 at radii $\leq$ 100 AU (i.e., nearly 5 times larger than the initial value). 
Therefore, S+12 detected flux transport within the Keplerian disk, at least in some of the disks formed. They did not consider that this could be due to transport arising from reconnection diffusion, because at these scales the velocity structures are already well ordered in their models. Thus S+12 concluded that numerical diffusion was the possible source of flux loss.
In this letter we put this conclusion to scrutiny and argue that the increase seen in S+12 is $real$ and due to reconnection diffusion in agreement with both theoretical expectation and the results of SGL12.

\section[]{Comparison between S+12 and SGL12 models}
Now, let us go back to the SGL12 results.  Focusing  on the formation of a Keplerian disk in   a turbulent cloud core with a single sink,  SGL12 clearly found flux loss during the process of the disk build up, as indicated from the analysis of their Figures 1 to 3.  In order to benchmark their results, SGL12 also  performed simulations of  non-turbulent  hydrodynamic as well as ideal MHD, and highly resistive MHD models.\footnote{In the later case, an artificially high resistivity about 3 orders of magnitude larger than the Ohmic resistivity was intentionally considered to allow comparison of the magnetic flux loss in this model with the turbulent model.}  They found that the ideal MHD model is unable to produce a rotationally supported disk due to the magnetic flux excess that accumulates in the central regions, while the MHD model with artificially enhanced resistivity produces a nearly-Keplerian disk with dimension, radial and rotational velocities,  and mass similar to the pure hydrodynamical model, and the turbulent model also produces a nearly-Keplerian disk, but less massive and  smaller  (r $\simeq$  100 AU), in agreement with the observations. 
For illustration, Figure A.1 of the Appendix, reproduces the  diagrams  as obtained in SGL12, i.e., radial profiles for   the rotational velocity, the  radial velocity, the mass and the magnetic field for the set of models described above, but  considering two different resolutions for the turbulent model (a 256$^3$ resolution, as in SGL12, and a 512$^3$ resolution model). As we see, both resolutions produce similar results (see more details in the Appendix).
In addition,  Figure 1 depicts three-dimensional diagrams of three snapshots of the turbulent model computed by SGL12.

 \begin{figure*}
 \includegraphics[width=158mm]{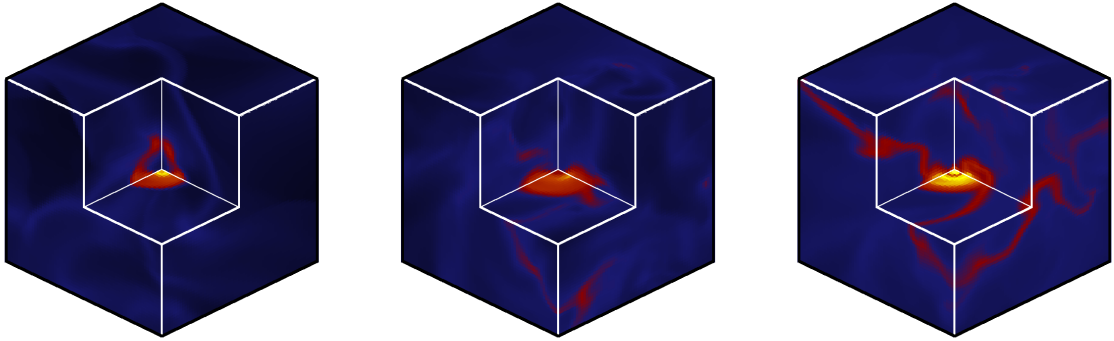}
 \caption{Three-dimensional diagrams of snapshots of the density distribution for the $turbulent$ model of  disk formation in the rotating, magnetized  cloud core computed by SGL12.  From left to right: t = 10.000 yr; 20.000 yr; and 30.000 yr. The side of the external cubes is 1000 AU (for more details see SGL12).}
 \label{fig1}
\end{figure*}

Considering the three MHD disk formation models investigated by SGL12 (i.e.,
an ideal collapsing cloud core with no turbulence, a highly resistive core with no turbulence, and an ideal turbulent core), Figure 2 shows the time evolution of the gas mass, the average magnetic flux and the mean mass-to-magnetic flux ratio, $\mu$, which was calculated employing the same equation (1) of S+12, for these three models. These quantities were computed within a sphere surrounding the central region for
 three different radii: a large one (r=1000 AU), which encompasses the large scale envelope where the disc is build up (similarly as in S+12), an intermediate (r=500 AU), and a small one (r=100 AU) which corresponds to the region where the disc is later formed.

 Figure 2 shows that the turbulent model of SGL12 starts with an average 
  $\mu \simeq 0.2$ and finishes with $\mu \simeq 0.5$ within $r= 1000$ AU
(see Figure 2, bottom right panel). Therefore, as in S+12, this result suggests no significant variations in $\mu$. Besides, these values  reveal no significant changes with respect to the ideal MHD model either. However, the values of  both, the turbulent  and the ideal MHD model at this scale, are also comparable to those of the resistive model where we clearly know that $there$ $is$ large magnetic flux loss.

 \begin{figure*}
 \includegraphics[scale=0.45]{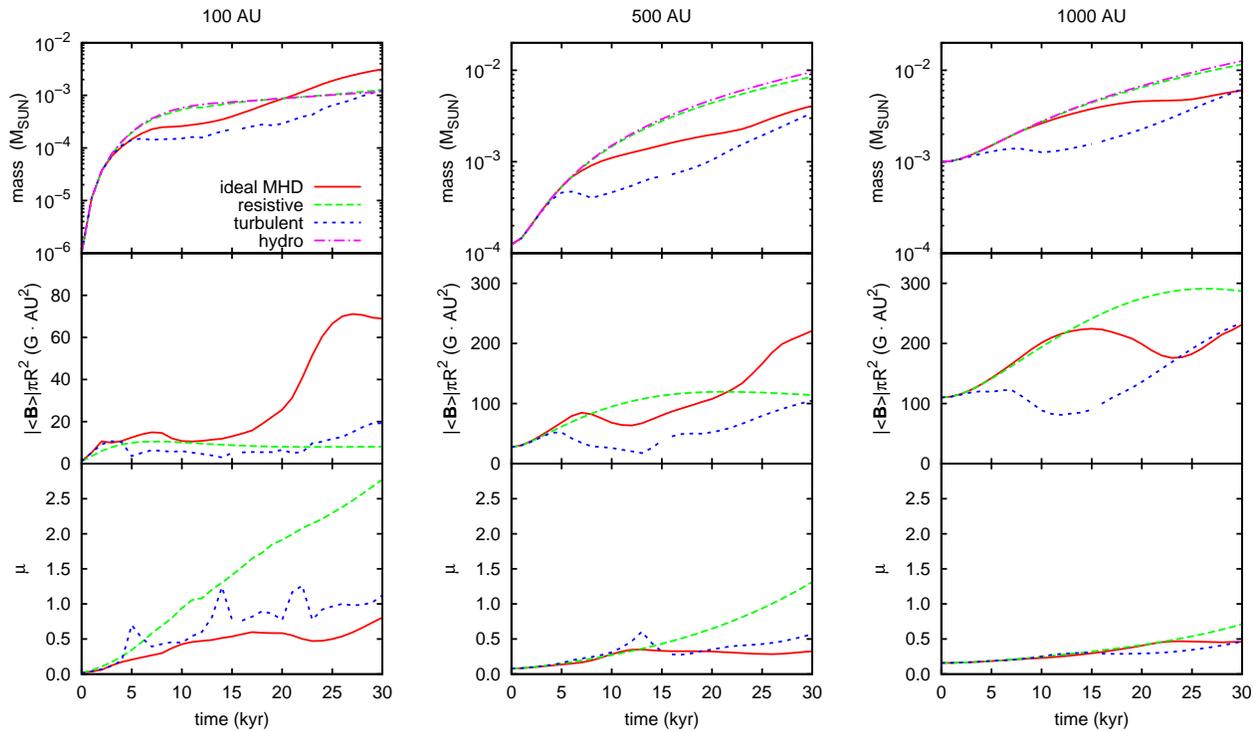}
 \caption{Disk formation in  the rotating, magnetized  cloud cores analysed by SGL12. Three cases are compared: an ideal MHD system, a resistive MHD system, and an ideal turbulent MHD system.  Right row  panels depict the time evolution of the total mass (gas + accreted gas onto the central sink) within a sphere of r=1000 AU (top panel),  the magnetic flux (middle panel), and the mass-to-flux ratio normalized by the critical value averaged  within r= 1000 AU (bottom panel). Left row  panels depict the same quantities for  r=100 AU, i.e., the inner sphere that involves only the region where the disk is build up as time evolves. 
 Middle row panels show the same quantities for the intermediate radius r=500 AU (see also Figure 1). 
 We note that the little bumps seen on the magnetic flux and $\mu$ diagrams for r=100 AU are due to  fluctuations of the turbulence whose injection scale ($\sim 1000$ AU) is much larger than the disk scale. }
\label{fig2}
\end{figure*}

How to interpret these results then? When averaging over the whole sphere of
radius $r= 1000$ AU around the disk/system, the real value of $\mu$ at the small disk scales (r $\le$ 100 AU) is hindered by the computed overall values in the envelope. Therefore, it is not enough to compute this average value to conclude that there is no flux loss in the process of the disk build up.

As we decrease the radius of the sphere at which the average $\mu$ is
computed, we clearly  see that the magnetic flux of the turbulent model becomes  comparable  to that of the resistive model (see middle panels of Figure 2), specially at the scale of the Keplerian disk build up ($r \simeq$ 100 AU) and, in consequence,  there is an increase of $\mu$ with time in the turbulent model with respect to the ideal MHD model.  The final value of $\mu \simeq 1$ within the disk region, therefore, nearly 5 times larger than the initial value in  the whole  cloud.

Thus, similarly to S+12, there is no significant variations of $\mu$ with time when considering its average over the whole sphere that contains both the turbulent envelope and the disk/sink. But,  in the final state the resulting value of $\mu$ within the disk is larger than the initial value in the cloud. A similar trend  is also found for the non-turbulent $resistive$ MHD model, where the imposed explicit artificial resistivity  leads to magnetic flux loss which in turn allows the build up of the Keplerian disk. These results clearly indicate that a nearly constant value of the average value of $\mu$ with time over the whole  disk$+$envelope system is not a powerful diagnostic to conclude that there is no significant magnetic flux transport in protostellar disk formation (as suggested by S+12).

When examining  the ideal MHD model, there is one important  point to  remark.
$\mu$, which should be expected to be
constant with time in an isolated system, is also  slightly growing within $r \simeq 100$ AU in this model.\footnote{At the larger radii this variation is hindered by averaging over larger scales as discussed.} 
This is due to the adopted  open boundaries in the system and to the volume averaging of the magnetic flux.
 To understand this behaviour, we must inspect the time evolution of both the mass and the average magnetic flux in the system which are shown in the top and middle  diagrams of Figure 2, respectively. 
Actually,   in all models  the total mass
(envelope/disk plus accreted gas into the sink)  increases with time due to a  continuous mass aggregation to the system entering through the open boundaries.
 Also, there is a  growing of the magnetic flux with time  which is at least in  part due to a continuous injection of magnetic field lines into the system through the open boundaries. Both effects, i.e., the increase in mass and magnetic flux could compensate each other and produce a nearly constant $\mu$ with time. However, there is another effect to be noticed. 
The magnetic flux   in the middle diagrams of Figure 2 was not computed  within a $comoving$ (accreting) volume with the gas, but at a fixed sphere radius.  If we had    followed  a fixed amount of  accreting   gas with time then,  we would have obtained  a constant number of magnetic field lines and therefore, a constant magnetic flux within this comovimg  volume in the ideal MHD case. This is in practice   very difficult  to compute from the simulations  because of  the complex geometry of the turbulent magnetic field lines. However, the key point here is not to obtain the exact value of the magnetic flux for the ideal MHD or the other models in a comoving volume, but to realize that  at the scale of the disk  build up  ($\sim$100 AU), the magnetic flux of the turbulent MHD model, which is initially comparable to that of the ideal MHD model,  decreased  to a value similar to that of the resistive model at the time that the disk has formed ($\sim$25,000  to 30,000 yr), as indicated by the middle left panel of Figure 2. This is a clear indication of the removal of magnetic flux  from the  disk  build up region to its surrounds in the turbulent model.

To  help to better clarify the analyses above, we have also plotted in Figure 3   $\mu$  as a function of the mass for the three different regions considered in Figure 2.  Each  $\mu(M)$ in Figure 3  has been normalized by its initial value: 
\begin{equation}
\mu_{0}(M) = \left[ \frac{M}{B_{0} \pi R^{2}_0(M)} \right] / \left[ 0.13/\sqrt{G} \right], 
\end{equation}
where $B_{0}$ is the initial value of the magnetic field and $R_{0}(M)$ is the initial radius of the sphere containing the mass $M$). These diagrams provide a way to evaluate  $\mu$ in comoving parcels with the gas.  
Inside 100 AU, Figure 3 shows that in the turbulent model $\mu(M)$ is larger  than in the ideal MHD model and comparable to the resistive model at the largest masses. This indicates a smaller amount of magnetic flux  in the turbulent and resistive models inside the disk region. As we go to the larger radii, $\mu$ becomes more and more comparable in the three models, in consistency with the results of Figure 2.

\begin{figure*}
 \includegraphics[scale=0.45]{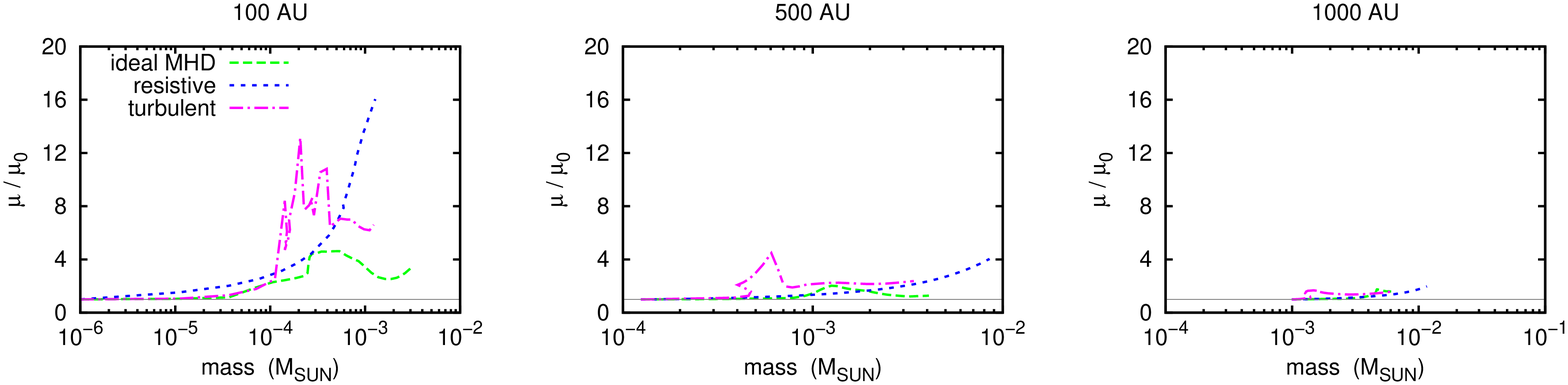}
 \caption{Mass-to-magnetic flux ratio $\mu$, normalized by its initial value $\mu_{0}(M)$, plotted against the mass, for (i) $r=100$ AU (left), (ii) $r=500$ AU (middle), and (iii) $r=1000$ AU (right). $\mu_0(M)$ is the value of $\mu$ for the initial mass $M$: 
$\mu_{0}(M) = \left\lbrace M / [ B_{0} \pi R^{2}_0(M) ] \right\rbrace / \left\lbrace 0.13/\sqrt{G} \right\rbrace$, where $B_{0}$ is the initial value of the magnetic field and $R_{0}(M)$ is the initial radius of the sphere containing the mass $M$. The initial conditions are  the same as  in Figure 2. 
}
 \label{fig3}
\end{figure*}

Therefore,  based on the results above, we conclude that the flux loss (and the increase of $\mu $ within the disk) in SGL12 turbulent models is REAL and is due to the action of
reconnection diffusion, as discussed in detail in SGL12 (see also Santos-Lima et al. 2010, Lazarian 2011, de Gouveia Dal Pino et al. 2012, Lazarian et al. 2012, Le\~ao et al. 2012). 

We must note that  the flux transport  by reconnection diffusion is faster where turbulence is stronger and faster. This is a fundamental prediction from  LV99 fast reconnection theory which was numerically  tested in high resolution simulations of cloud collapse in Santos-Lima et al. (2010; see also the Appendix).  In the SGL12 turbulent simulations,  while the disk is built up by the accretion of the turbulent gas in the envelope that surrounds the sink, reconnection diffusion is fast and causes magnetic flux loss at the same time that it allows the turbulent shear to build up a Keplerian profile  in this \textit{collapsing}  material. This means that  the material that formed the Keplerian disk out of the accretion of the turbulent envelope {\it has already lost magnetic flux}  when it reaches its final state and that is why the final value of $\mu$  is much larger within the disk radius ($\le$ 100 AU).
In other words, the mass-to-flux ratio increase that is detected in the final
Keplerian disk is due to removal of magnetic flux from the highly turbulent
envelope material while this material was accreting and building up the disk,
i.e., \textit{before} the final state. After the Keplerian disk is formed (in r $\le$ 100 AU), the operation of reconnection diffusion inside this region decreases
because turbulent structures are smaller and slower there. Fortunately,
in terms of magnetic breaking, high values of reconnection diffusion are no
longer needed because the magnetic field flux excess has been already
removed during the accreting phase and disk build up.

As a result, the argument given by S+12 that reconnection diffusion could not
explain the increase of  $\mu$ in their tests within the disk scales because the fluid is no longer turbulent there, is not correct.

We  have also plotted the magnetic field $B$ versus the density $\rho$ in different times as in S+12 and found variations which are significant as we decrease the radius where the average of B and $\rho$ are computed, in consistency with the results of Figures 2 and 3 and the discussion above.
 Figure 4  shows these plots for 30,000 yr within spheres of radii equal 100, 500, and 1000 AU. At the 1000 AU scale, both the ideal and the turbulent MHD models are comparable and follow  approximately  the $B \propto \rho^{0.5}$ trend,  as in S+12. However, as we go to the smaller scales and specially to the 100 AU scale, the two models clearly loose this correlation, similar to the resistive model in all scales. This effect in the resistive  model is clearly a natural consequence of the diffusion of the magnetic field from the inner denser regions to the less dense envelope regions.  The turbulent model tends to follow the same trend:  we note  that for a given density, the magnetic field is smaller in the resistive model than in the turbulent model which  in turn, has a smaller magnetic field than in the ideal model (see left panel in Figure 4), in consistence with the previous results. In the case of the MHD model, the nearly constant magnetic field with density   at the 100 AU scale is due to the effect of the geometry.  At this scale, the built up disk dominates, but the averaging is performed  over the whole sphere that encompasses the region. This includes also the very light material above and below the disk which has magnetic field intensities as large as those of the high density material in the disk. (The same effect explains also the larger magnetic field intensities in the low density tail at the 500 and 1000 AU scale diagrams $-$ middle and right panels,  specially for the ideal MHD model.) The geometry at 100 AU  obviously also affects the turbulent model in the same way, however we have found from the simulations that in this case the amount of low density gas carrying high intensity magnetic field below and above the disk is  smaller than in the ideal MHD case. This is because in this case the loss of the $B-\rho$ correlation at the 100 AU scale is also  affected by the diffusion of the magnetic field as in the resistive model, in consistence with the analyses of Figures 2 and 3. 

 \begin{figure*}
 \includegraphics[scale=0.45]{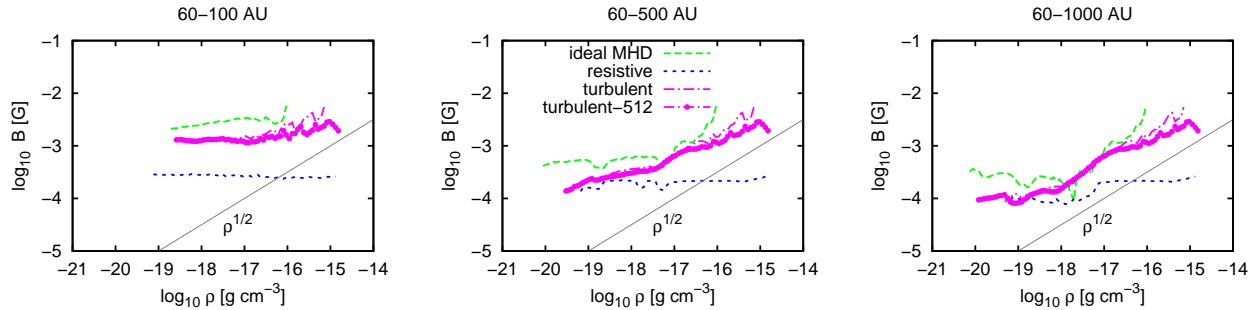}
 \caption{Mean magnetic intensity as a function of bins of  density, calculated for the models analysed in SGL12  at $t=30$ kyr. The statistical analysis was taken inside spheres of radius of $100$ AU (left), $500$ AU (middle), and $1000$ AU (right). Cells inside the sink zone  (i.e., radius smaller than $60$ AU) were excluded from this analysis. For comparison, we have also included the results for the turbulent model  \textit{turbulent-512} which was simulated with a resolution twice as large as the model \textit{turbulent-256} presented in SGL12 (see also the Appendix). }
 \label{fig4}
\end{figure*}

Although the initial conditions in SGL12  are different from those in S+12, the build up of the Keplerian disks by the accretion of the turbulent envelope around the sinks is quite similar to the setup in SGL12, so that we can perform at least qualitative comparisons between the results.
 In particular, both models consider initially supercritical cores\footnote{Supercritical cores have a mass-to-magnetic flux ratio which is larger  than the critical value at  which the magnetic field force balances  the gravitational force.  Subcritical cores, satisfy  the opposite condition.}, i.e., initial $\mu$ larger than unity. We remember  that  in Figures 2 and 3 only the values of $\mu$ corresponding to the accreted gas  mass were plotted in order to allow  an easier track of  the mass and magnetic flux evolution of the disk and envelope material. Nonetheless,  the total $\mu$ in the SGL12 models are larger than  unity when including the sink, as in S+12 models (see SGL12). Another  important parameter in this analysis of magnetic flux transport is the initial ratio between the thermal and magnetic pressure of the gas, $\beta$, which is also similar in both models ($\beta \sim 0.1$ in the center of the cloud in   the S+12 models, while  $\beta \sim 0.6$ in the  whole core of the  SGL12 models).  
As a matter of fact,  SGL12  chose  an initial value of $\beta$  smaller than unity  in the core in order  to show that 
 reconnection diffusion could be able to remove the magnetic flux excess even from an initially magnetically dominated   gas and thus solve 
the magnetic braking problem. In the case of the S+12 models, it is possible that due to their setup,  sink regions far from the center of the cloud 
do not have this constraint on $\beta$. In this case,   such regions  might not  require, in principle,   
removal of magnetic flux to allow the formation of a rotating disk   and thus this initial condition would naturally avoid  the magnetic braking problem. 
However, when ingredients such as rotation and turbulence are introduced, the accreting  history in the different sinks within this cloud may change completely depending on the relative strength of the turbulence and magnetic field.  In this sense, the formation of a Keplerian disk is very sensitive to the $local$  
conditions of the region around the sink (rather than the global initial conditions of the entire cloud) and therefore, the track of the detailed evolution of the magnetic flux around  each sink region at the scale of the disk build up would be  required in order to evaluate the real evolution of $\mu$ around each sink. Such an analysis is missing in S+12 work.

Based on the discussion in the previous paragraphs, it is  natural to assume that the increase in $\mu$
detected in some of the S+12 models within the disk radius is $real$, as in
SGL12 model. This increase could be simply an evidence that flux loss was very efficient during the disk build up in the turbulent envelope around this sink. The fact that they find a final $\mu $ in the disk which is much larger than the initial value in the cloud suggests that even if numerical resistivity is operating in the inner regions,  a
substantial magnetic flux excess was removed by turbulent reconnection diffusion when the disk was still forming from the accreting envelope.

Regarding the potential role of the numerical resistivity, we can make some quantitative estimates.  
The relevant scales for the reconnection diffusion to be operative are the turbulent scales, from the injection  to the dissipation scale, i.e., within the inertial range scales of the turbulence which are  larger than the  numerical viscous scale. 
In the simulations of SGL12 with a resolution of $256^3$ this scale is  approximately of 8 cells. To evaluate the relative role of the numerical dissipation on the evolution of the magnetic flux at scales near the dissipation range, we can compare the advection and the diffusion terms of the  magnetic field induction equation at a given scale. Considering the magnetic flux variations of the SGL12 turbulent model within a $100$ AU scale at $t = 30$ kyr, we find that the ratio between these two terms, which gives the magnetic Reynolds number, is  $R_m = L V/\eta_{Num} \sim 75$,  where  $L=100$ AU and we have considered the radial  infall velocity  as a characteristic velocity of the system in this region, $V \approx 0.5$ km/s (see Figure A.1). The approximate numerical resistivity for the employed resolution in SGL12 is $\eta_{Num} \sim 10^{18} cm^{2} s^{-1}$.  Therefore, although present, the numerical dissipation was not the dominant ingredient driving the change of the magnetic flux inside $100$ AU in the SGL12 turbulent model. 
Examining the case of the S+12 models with increase in  $\mu$, considering that their magnetic Reynolds number  must be even larger at 100 AU  (i.e.,  the numerical viscosity  is even smaller)  due to  their higher  resolution, then in their case there should be no  significant magnetic flux removal either at 100 AU due to numerical resistivity  because of the same arguments above (see also the Appendix).  

Since the {\bf S+12} authors do not provide the details of the magnetic field,  turbulence, and density intensity within their Keplerian disks, it is hard to argue whether there was some significant flux loss or not in the other cases that they investigated where no increase was detected in the averaged $\mu$ over a large volume. It is also possible that some of these disks developed in regions where the local magnetic fields were not strong enough to cause magnetic braking and prevent the growth of the Keplerian disk. In these cases, even if flux loss by reconnection diffusion is occurring it would be undetectable.

\section{ Discussion and Conclusions}
In the previous section, we have demonstrated  that an analysis which is based solely on the computation of the average value of the mass-to-magnetic flux ratio ($\mu$)  over the whole envelope that surrounds a newly formed Keplerian disk, is not adequate  to conclude that there is no significant magnetic flux loss in simulations of disk formation.  This averaging masks significant real increases of $\mu$ in the inner regions where the disk is build up out of the turbulent envelope material that is accreting.

Actually, we have demonstrated that this is what happens on the build up of the Keplerian disk both, in the turbulent and  in the resistive models of SGL12, where magnetic flux loss has been detected. While the average $\mu$  computed over the large scale envelope/disk does remain nearly constant with time, the value of $\mu$ inside the formed Keplerian disk is 5 times larger than the initial value in the cloud core. Similarly, some of the disks formed in the turbulent S+12 models also revealed a value of  $\mu$  5 times larger within the disk than the initial cloud value, while the average $\mu$ over the whole envelope surrounding the disk was nearly constant with time.

In SGL12 models it has been found that the reduction of the magnetic braking efficiency during the building of the disk was due to the action of reconnection diffusion.  The latter process is based on the LV99 theory, as remarked, which in turn, has been tested with high resolution numerical simulations \citep{kowal_etal_2009, kowal_etal_2012a} as well as, by Lagrangian
analysis of turbulent MHD fluids \citep{eyink_etal_2011}. In addition, reconnection diffusion has been successfully tested numerically for the first time in \citet{santos-lima_etal_2010} 
with very high resolution simulations 
which, taken together, allows us to claim that the process must be occurring in all magnetized turbulent environments, including regions of accretion disk formation (see also Le\~ao et al. 2012, de Gouveia Dal Pino et al. 2012).
Therefore, we suggest that the increase in $\mu$ found in S+12's  disk is real and, as in SGL12, is caused by  reconnection diffusion rather than numerical effects. While the shear flow generated by the turbulent  motions in the surroundings of the disk (which carry large amounts of angular momentum) allows the build up of the rotationally supported disk, it also removes the magnetic flux excess due to fast turbulent reconnection, which otherwise may prevent the formation of the disk. During the build up of these disks out of turbulent envelopes around the sinks embedded in a massive cloud core (as in S+12 models), some envelopes may have strong local magnetic support (low local $\mu$) which may prevent the Keplerian disk formation unless magnetic flux is removed, and some not. In the former case, we argue that reconnection diffusion is reducing the effect of the magnetic braking. In the latter case, even in the presence of magnetic flux loss induced by reconnection diffusion in the turbulent flow, its
effect is marginal and therefore difficult to detect, because the magnetic field is dynamically unimportant.

Naturally, both SGL12 and S+12 models suffer from numerical effects at small scales. However, we have provided quantitative arguments that evidence that numerical resistivity should have a minor effect in both cases. 
Besides, it is important to emphasize  that reconnection diffusion is not just an empirically explored phenomenon, as remarked above. In particular, it has been shown that the LV99 theory does not depend on the microphysics of local reconnection events \citep{kowal_etal_2009, kowal_etal_2012a}. 
Therefore,  reconnection diffusion should be represented correctly  by numerical simulations at the scales where the turbulence is  not damped (see also the Appendix and  Lazarian 2011, Santos-Lima et al. 2010, de Gouveia Dal Pino et al. 2012, Le\~ao et al. 2012). Since this condition is satisfied at the scales investigated in  the numerical simulations here discussed, reconnection diffusion is expected to be represented correctly in these simulations.

\section*{Acknowledgments}

R.S.L acknowledges support from the Brazilian agency FAPESP (2007/04551-0) and E.M.G.D.P from FAPESP (2006/50654-3) and CNPq (300083/94-7).
AL acknowledges the support the NSF grant AST 1212096, the Vilas Associate Award as well as the NSF Center for
Magnetic Self-Organization in Laboratory and Astrophysical Plasmas. The calculations presented here were performed in the supercomputer Alfa-Crucis of the Astrophysical Informatics Laboratory LAi of IAG, Astronomy Department, University of S{\~a}o Paulo (funded by FAPESP: 2009/54006-4).

\appendix

\section[]{Effects of numerical resolution on the turbulent model}

Santos-Lima et al. (2010) have performed a rigorous numerical test of the role of turbulent magnetic reconnection diffusion on the transport of magnetic flux in diffuse cylindrical clouds, considering periodic boundaries and different numerical resolutions between $128^3$ and $512^3$. They  found very similar results for all  resolutions which revealed the importance of the  diffusion mechanism above to remove magnetic flux from the inner denser to the outer less dense regions of the clouds. 
In particular, they  demonstrated that the higher the value of the ratio between thermal and magnetic pressure,  $\beta$ (smaller B) the more efficient the flux removal due to turbulent reconnection diffusion is. 
Recently, similar studies with self-gravitating spherical magnetized  clouds have  confirmed this trend as well (Le\~ao et al. 2012, de Gouveia Dal Pino et al. 2012).  In particular, they have demonstrated  that turbulent reconnection diffusion of the magnetic flux is very effective and may allow the  transformation of  initially subcritical into  supercritical cores.

The turbulent reconnection diffusion theory is based on the fact that in the presence of turbulence, magnetic reconnection becomes fast and independent of the Ohmic resistivity (at the scales where the magnetic Reynolds number $R_{m}$ is $\gtrsim 1$). This is because  there is an increase  of the number of magnetic reconnection events boosted by the turbulence, in according with Lazarian \& Vishniac (1999) reconnection theory (which  has been already tested  numerically in Kowal et al. 2009, 2012). However, a common distrust in turbulent numerical simulations is that turbulence could be enhancing the numerical resistivity itself. This is unjustified, as proved  by the careful numerical analysis performed in Kowal et al. (2009, 2012) and in Santos-Lima et al. (2010). If this were true, the turbulent reconnection rate would reduce whenever the resolution of the numerical experiment were increased, which is not the case. Besides, the turbulent reconnection diffusion coefficient, which is of the order of the Richardson hydrodynamical diffusion coefficient ($\eta_{turb} \simeq L V_{turb}$ for  super-Alfv\'enic turbulence; see e.g., Lazarian 2011),  is much larger than the numerical (or the Ohmic) diffusion   coefficient at  scales larger than the dissipation scales of the turbulence, so that the effects of turbulent reconnection diffusion on the magnetic flux transport are  dominant over numerical diffusion at the relevant scales of the system.

Nonetheless, in order to provide more quantitative tests about the reliability of the SGL12 models, we  also have run a similar turbulent model to  that in SGL12, but with a resolution twice as large, which is named \textit{turbulent-512}. We also included in this model an explicit small Ohmic resistivity ($\eta_{Ohm} = 10^{17} cm^{2} s^{-1}$) in order to speed up the numerical computation (this is however, comparable to the numerical viscosity for this resolution and much smaller than the turbulent diffusion coefficient, so that it does not influence the physical results of the problem). The computational domain  in this model is cubic with $4000$ AU of side, and the sink accretion radius is half of the value for the models in SGL12 ($\approx 30$ AU). In Figure A.1, we present the radial profiles for the mean values of the radial and azimuthal velocities, the disk (+ envelope) mass,  and the mean vertical magnetic field at $t=30$ kyr for this model, which are compared with those of   the SGL12 models.

 \begin{figure*}
 \includegraphics[scale=0.6]{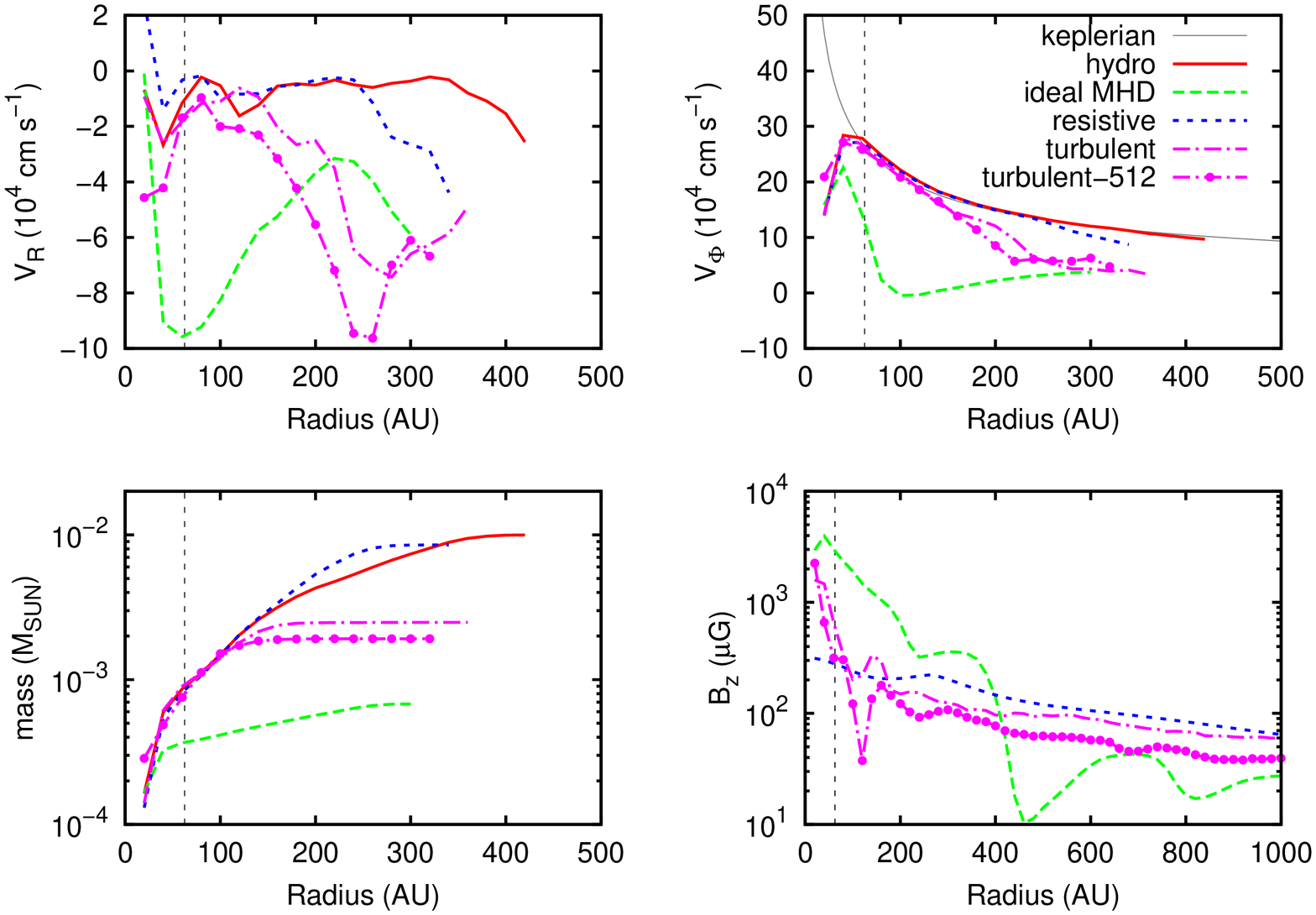}
   \caption{Comparison between the radial profiles of the high resolution turbulent model \textit{turbulent-512} with the models presented in SGL12 (for which the resolution is $256^3$). Top left: radial velocity $v_{R}$. Top right:  rotational velocity $v_{\Phi}$. Bottom left: inner disk mass. Bottom right: vertical magnetic field $B_{z}$. The numerical data are taken at time $t\approx 0.03$ Myr. The velocities were averaged inside cylinders centered in the protostar with height $h=400$ AU and thickness $dr=20$ AU.  The  magnetic field values were also  averaged  inside equatorial rings centered in the protostar. 
The vertical lines indicate the radius of the sink accretion  zone for all models except  \textit{turbulent-512} for which the the sink radius of the  accretion zone is half of that value.}
 \label{fig5}
\end{figure*}

The overall behaviour of these models is described in detail in SGL12. Here, we only  emphasize the main features. We  clearly see that the results of the turbulent models for both resolutions are similar. A disk with a Keplerian velocity profile is also build up in the high resolution model  within the same radius ($\approx 100$ AU) as in  the smaller resolution turbulent model of SGL12. The mass of the disk inside this radius is  also identical in both models and slightly smaller for larger radii (in the envelope) in the higher resolution model. The vertical magnetic field is also slightly smaller for radii larger than $500$ AU, but similar to the lower resolution model in  the inner regions (except for some fluctuations due to the different sizes of the  sinks, but which are not relevant for the present analysis. The infall velocity is generally smaller in the higher resolution model.  Nonetheless, in this model,  a slightly thinner disk develops and  if this velocity is averaged only over the  higher density gas concentrated at smaller heights around the disk, then the infall velocity profile becomes  very similar  with that of the smaller resolution model. This also has to do with  the fact that the turbulence in the model with lower resolution decays slightly faster, so that  at the period of time considered in  Figure A.1, the model with lower resolution has reached already a more relaxed, non turbulent state. 
The similarity between the results of both turbulent models indicates that the lower resolution model of SGL12  is reliable and therefore, can be employed in the analysis presented in this work.

\label{lastpage}

\end{document}